\documentclass[aps,prb,twocolumn,showpacs,superscriptaddress,
preprintnumbers,amsmath,amssymb]{revtex4}

\usepackage{graphicx}
\usepackage{dcolumn}
\usepackage{bm}
\usepackage{ifthen}
\usepackage{booktabs}
\def\mc{\multicolumn}

\begin{document}

\title{Electronic structure of self-assembled InAs/InP quantum dots: 
A Comparison with self-assembled InAs/GaAs quantum dots}
\author{Gong Ming}
\affiliation{Key Laboratory of Quantum Information, 
University of Science and Technology
of China, Hefei, 230026, People's Republic of China}
\author{ Kaimin Duan}
\affiliation{Key Laboratory of Quantum Information, 
University of Science and Technology
of China, Hefei, 230026, People's Republic of China}
\author{ Chuan-Feng Li}
\affiliation{Key Laboratory of Quantum Information, 
University of Science and Technology
of China, Hefei, 230026, People's Republic of China}
\author{Rita Magri}
\affiliation{Cnism-CNR and Dipartimento di Fisica, 
Universit\"a di Modena e Reggio Emilia, Via Campi 213/A, 41100
Modena, Italy}
\author{Gustavo A. Narvaez}
\affiliation{Amin, Turocy \& Calvin, LLP, Cleveland, OH 44114, USA}
\author{Lixin He \footnote{corresponding author: helx@ustc.edu.cn} }
\affiliation{Key Laboratory of Quantum Information, 
University of Science and Technology
of China, Hefei, 230026, People's Republic of China}
\date{\today }

\begin{abstract}

We investigate the electronic structure of the 
InAs/InP quantum dots using an
atomistic pseudopotential method
and compare them to those of the
InAs/GaAs QDs. 
We show that even though the InAs/InP and InAs/GaAs 
dots have the same dot material,
their electronic structure differ significantly 
in certain aspects, especially for holes: 
(i) The hole levels have a much larger 
energy spacing in the InAs/InP dots than
in the InAs/GaAs dots of corresponding size. 
(ii) Furthermore, in contrast with the InAs/GaAs dots, 
where the sizeable hole $p$, $d$ intra-shell level splitting 
smashes the energy level shell structure, the InAs/InP QDs
have a well defined energy level shell structure
with small $p$, $d$ level splitting, for holes.
(iii) The fundamental exciton energies of the InAs/InP dots
are calculated to be around 0.8 eV
($\sim$ 1.55 $\mu$m), about 200 meV lower than those of typical InAs/GaAs
QDs, mainly due to the smaller lattice mismatch in the InAs/InP dots. 
(iii) The widths of the exciton $P$ shell and $D$ shell
are much narrower in the InAs/InP dots than in the InAs/GaAs dots.
(iv) The InAs/GaAs and InAs/InP dots 
have a reversed light polarization anisotropy
along the [100] and [1$\bar{1}$0] directions.

\end{abstract}


\maketitle


\bibliographystyle{prsty} 

\section{Introduction}

Self-assembled semiconductor quantum dots (QDs) have attracted a large
interest because the discrete and isolated energy levels due 
to the 3D confinement can be utilized in high-efficiency 
and low-threshold current lasers,\cite{caroff05, huffaker98, park99, shchekin02}
single photon emitters\cite{michler00, benson00} and  
in quantum computing applications (qubits).
\cite{loss00, bayer01}
1.3 $\mu$m optical devices have been achieved 
using InAs/GaAs quantum dots,\cite{huffaker98, huffaker98b, park99}
however, it is difficult to push further the InAs/GaAs QD devices
to work at the more desirable $\sim$1.55$\mu$m telecommunication wavelength 
because the large compressive strain (7\%) in the InAs/GaAs dots
enlarges the conduction-valence energy gap of 
InAs too much for this purpose.
Intuitively, a InAs dot embedded in a less lattice mismatched 
host material, e.g. InP, should solve the problem.
Indeed, recently there have been reports on 1.55 $\mu$m InAs/InP QDs
lasers with a high-gain and a low-threshold  current,\cite{caroff05, allen02} 
which opens up the possibility to integrate quantum dot material into an 
optical cavity\cite{dalucn04} and hence to reliably fabricate a single 
photon source in the telecommunications 
wavelength range.\cite{michler00, benson00}

Despite their importance, there are only few studies
on the InAs/InP system, both theoretically \cite{sheng05b, cornet06a}
and experimentally \cite{pettersson00, chithrani04, cade06, dupuy06,ujihara06, gendry04}
compared to the well studied InAs/GaAs QDs. 
Although the InAs/InP dots and the InAs/GaAs dots have the same dot material,
they differ in three aspects, 
(i)  InAs/InP  has a  much smaller lattice mismatch (3\%) than 
InAs/GaAs (7\%).
(ii) The InAs/InP dots have a less confining potential for electrons,
but a stronger confinement for holes than the InAs/GaAs dots.
(iii) The InAs/InP dots share the same cation (In), while the
InAs/GaAs QDs share the same anion (As) at the interface.  
These differences may lead to different electronic and optical properties of 
the two systems.
The continuum theories, such as those based on the effective mass approximation (EMA) and 
multi-bands  ${\bf k}\cdot {\bf p}$ method, may in principle
capture the first two differences, however, to account for the third aspect,
one needs atomistic theories such as the empirical pseudopotential methods 
\cite{wang99b,williamson00} or the
tight-binding methods.\cite{sheng05b, seungwon01}
The empirical pseudopotential methods have been
successfully applied to various systems. \cite{williamson00,he05a} 
In this paper, we perform a comparative
study on the InAs/InP QDs and InAs/GaAs QDs using an atomistic
pseudopotential method. We find that there are significant
differences in the electronic structure between the two systems, including  
the single-particle energy levels and optical properties.
These differences, which have not yet been paid enough attention to, 
could be revealed in the future high-resolution optical
spectroscopy \cite{kuther98, hawrylak00} 
and charging experiments. \cite{drexler94,reuter05} 

The rest of the paper is organized as
follows. In Sec. \ref{sec:methods}, we introduce briefly the atomistic
pseudopotential method used in the calculations. 
In Sec. \ref{sec:strain},
we compare the strain profiles and the strain modified band-offsets for the
InAs/GaAs and InAs/InP QDs.
In Sec. \ref{sec:single-particle}, we compare  
the pseudopotential calculated electronic structure 
as well as the wavefunctions of the two dots.
We compare the excitonic transitions 
of the two dots in Sec. \ref{sec:excitons}
and conclude in Sec. \ref{sec:conclusion}.

\section{Methods}
\label{sec:methods}

The geometries of the QDs studied here are lens-shaped
InAs dots embedded in the host materials (GaAs or InP) matrices
containing 60 $\times$ 60$\times$ 60 8-atom unit cells. 
The dots are assumed to grow along the [001] direction. 
We performed the calculations on dots with 
base diameters $D$= 20, and 25 nm, 
and for each base diameter, we vary the dot heights $h$
from 2.5 nm to 5.5 nm, as the dot height is 
relatively easy to control in experiments.\cite{paranthoen01}
We first relax the dot+matrix system by minimizing the strain energy as 
a function of the coordinate $\{ {\bf
R}_{n,\alpha} \}$ of atom $\alpha$ at site $n$ for all atoms, 
using valence force field (VFF) methods.\cite{keating66,martin70} 
Once we have the atom positions,
we obtain the energy levels and wavefunctions
by solving the single-particle Schr\"{o}dinger equation, 
\begin{equation}
\left[ -{1 \over 2} \nabla^2 
+ V_{ps}({\bf r}) \right] \psi_i({\bf r})
=\epsilon_i \;\psi_i({\bf r}) \; ,
\label{eq:schrodinger}
\end{equation}
where $V_{ps}({\bf r}) = V_{SO}+ \sum_n\sum_{\alpha} v_{\alpha}({\bf r} 
- {\bf R}_{n,\alpha})$ is the superposition of
local screened atomic pseudopotential{\bf s} $v_{\alpha}({\bf r})$, and the total
(non-local) spin-orbit (SO) potential $V_{SO}$.
The screened pseudopotentials~\cite{williamson00}
are fitted to the physical important
properties of the materials, including the band energies at high-symmetry
points, effective masses, strained band offsets,
hydrostatic and biaxial deformation potentials of
individual band edges. 
The pseudopotentials of InAs/InP dots
are given in Appendix \ref{sec:EPM}, 
whereas the potentials for InAs/GaAs QDs are taken
from Ref.~\onlinecite{williamson00}.
The Schr\"{o}dinger Eq. (\ref{eq:schrodinger}) is solved 
by expanding the wavefunctions
$\psi_i$  as linear
combinations of bulk bands (LCBB)~\cite{wang99b} 
$\{\phi_{m,\tensor{\epsilon},\lambda}({\bf k})\}$  of 
band index $m$
and wave vector ${\bf k}$ of material $\lambda$ (= InAs, GaAs, InP), strained
uniformly to strain $\tensor{\epsilon}$. 
We use $m=2,3,4$ for the hole states,
and $m=5$ for electron states on a
6$\times$6$\times$16 k-mesh. We use  $\tensor{\epsilon}=0$ for the
matrix material, and an average $\tensor{\epsilon}$ value 
from VFF for the strained dot material (InAs).
It has been shown that the energy levels changes in InAs/GaAs
QDs due to the piezoelectric effects are quite small. \cite{bester05a}
Because the lattice mismatch in the InAs/InP QDs is only half
of that of the InAs/GaAs QDs, 
we expect that the  piezo-effect should be even smaller in the InAs/InP dots, 
and therefore is ignored in the present calculations.

The exciton energies are calculated using the configuration interaction (CI)
method following Ref. \onlinecite{franceschetti00}.

\begin{figure}
\begin{center}
\includegraphics[width=3in,angle=-90]{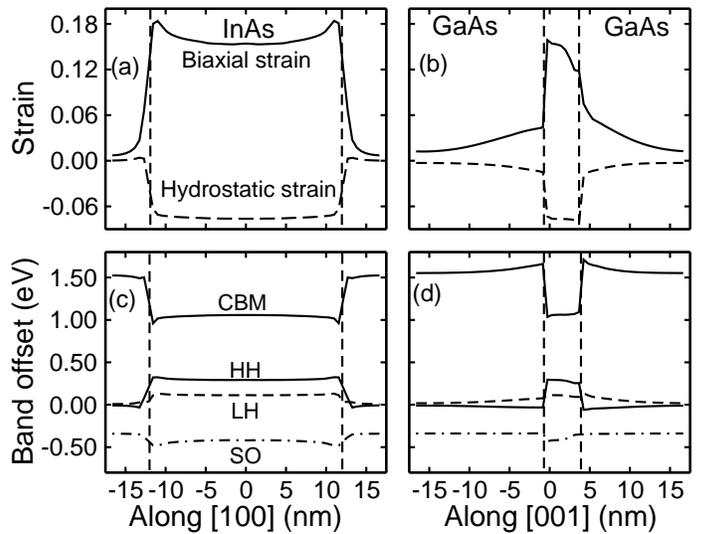}
\end{center}
\caption{Strain profiles and strain modified band-offsets for
lens-shaped InAs/GaAs QD ($D$=25 nm,
$h$=3.5 nm). The strain profiles are shown in (a)
along the [100] direction and in (b) along the [001] direction.
The strain modified band-offsets are shown in (c)
along the [100] direction and in (d) 
along the [001] direction for the CBM, HH, LH and SO bands.  
The reference energy is chosen to be the VBM of GaAs.}
\label{fig:strain_gaas}
\end{figure}

\begin{figure}
\begin{center}
\includegraphics[width=3in,angle=-90]{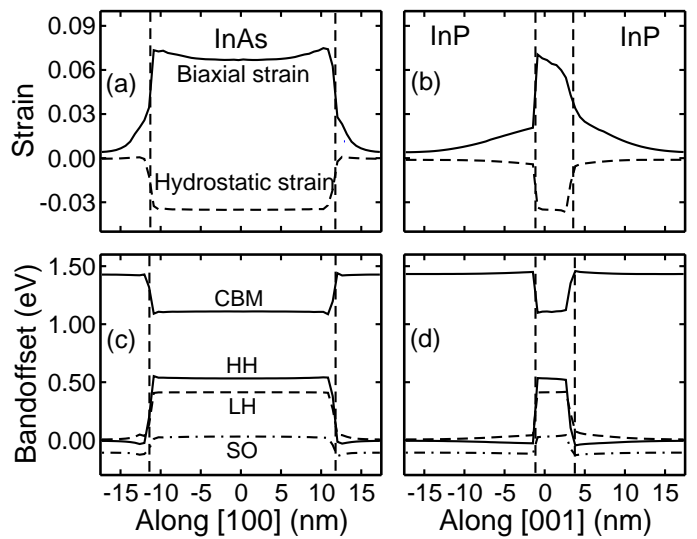}
\end{center}
\caption{Strain profiles and strain modified band-offsets for
lens-shaped InAs/InP QD ($D$=25 nm,
$h$=3.5 nm). The strain profiles are shown in (a)
along the [100] direction and in (b) along the [001] direction.
The strain modified band-offsets are shown in (c)
along the [100] direction and in (d) 
along the [001] direction for the CBM, HH, LH and SO bands.  
The reference energy is chosen to be the VBM of InP.}
\label{fig:strain_inp}
\end{figure}

\section{Strain profiles and strain modified bandoffsets}
\label{sec:strain}

We first compare the strain profiles in the InAs/GaAs and the InAs/InP QDs.
Figure \ref{fig:strain_gaas} (a), (b)  
depict the hydrostatic and biaxial strain along the [100] direction 
and the [001] direction respectively for the InAs/GaAs dot,
whereas Fig. \ref{fig:strain_inp} (a), (b)
depict the strain profiles for the InAs/InP quantum dots.
The strain is calculated for the lens-shaped QDs with $D$=25nm, and 
$h$=3.5nm. The hydrostatic(isotropic) strain is defined as
${\rm I}= {\rm Tr}(\epsilon)=\epsilon_{xx}+\epsilon_{yy}+\epsilon_{zz}$, 
reflecting the relative change of volume i.e., ${\rm I} \sim \Delta V/V$,
whereas the biaxial strain is defined as
${\rm B=}\sqrt{(\epsilon_{xx}-\epsilon_{yy})^2+
(\epsilon_{zz}-\epsilon_{xx})^2+(\epsilon_{yy}-\epsilon_{zz})^2}$.  
As we can see, the hydrostatic
strains of the InAs/InP and InAs/GaAs dots have very similar features:
both are almost constants inside the dots and 
decay rapidly to zero in the matrices.
The hydrostatic strain is negative in both dots, suggesting
that the InAs dot is compressed.
However, the hydrostatic strain in the InAs/InP dot 
is about half of that in the InAs/GaAs dot. 
The biaxial strain of both QDs also has similar
features,  decaying fast in the [100] and [010] directions outside
the dots,  it has long tails along the [001] direction.

The hydrostatic strain and the biaxial strain will shift the conduction band
minimum (CBM) and the valence band maximum (VBM), and therefore modify the
band-offsets of the QD. The biaxial strain will further 
split the heavy-hole (HH) and light-hole (LH) bands.
We analyze the strain-modified band-offsets via Pikus-Bir
model, \cite{pikus,wei94, he04a, pryor98c} using the local strain input 
from VFF calculations. 
This model, however, serves only as an illustration of the
strain effect and is not used in our actual
calculation of the single particle energy levels.
The strain modified band-offsets are illustrated in 
Fig. \ref{fig:strain_gaas} (c),(d) 
for InAs/GaAs quantum dot, and in
Fig. \ref{fig:strain_inp} (c),(d)
for InAs/InP quantum dot along the [100] direction
and [001] direction respectively.
The VBM of the unstrained host materials is set
to be zero as the reference energy.  
The band-offsets in the InAs/GaAs dot
are 480 meV for the electron and 280 meV for the hole, compared with
1050 meV for the electron and 46 meV for the hole in the unstrained system. 
\cite{vurgaftman01} The confinement potential for electrons is stronger 
than that for holes. 
In contrast, for the unstrained InAs/InP system, 
the band-offset\cite{vurgaftman01} is 580 meV for the electron  
and 420 meV for the hole, which change to 320 meV
and 530 meV respectively in the InAs/InP
QDs, due to the strain effects. The confinement potential for holes is  stronger
than that for electrons.\cite{pettersson00}
Therefore, the confinement potentials in the InAs/GaAs and the InAs/InP QDs are 
very different. While the confinement for electrons is weaker in the 
InAs/InP QDs, the confinement for holes is significantly stronger.
How the different confinement potentials lead to different electronic 
structure in the two dot systems will be discussed in Sec. \ref{sec:single-particle}.    
It was pointed out in Ref. \onlinecite{he04a} that in the tall InAs/GaAs QD, the
biaxial strain might develop hole traps that
localize holes at the interface of the QDs. This is unlikely to happen in
the InAs/InP QDs for two reasons. (i) The strain in the InAs/InP QDs is much
smaller than in the InAs/GaAs QDs. (ii) The band-offset for holes
in the InAs/InP QDs is much larger than in the InAs/GaAs QDs.

The biaxial strain splits the HH and LH bands in addition to shifting the VBM. 
HH is higher in energy than the LH band, 
i.e., $E_{hh} > E_{lh}$ inside both dots. The HH-LH splitting is
about 180 meV for the InAs/GaAs QD and 120 meV for the
InAs/InP QD.  Outside the QD, the heavy-light-hole
splitting changes sign, i.e., $E_{hh} < E_{lh}$. \cite{he04a}
It is also interesting to note that the SO band inside the InAs/GaAs quantum
dot is lower in energy than outside the QDs, while the opposite is true for the 
InAs/InP quantum dot, due to the large band-offsets between
InAs and InP for holes.

\section{Single-particle energy levels and wavefunctions}
\label{sec:single-particle}

In this section, we compare the pseudopotential calculated 
single-particle energy levels as well as the wavefunctions 
of the InAs/InP QDs 
to those of the InAs/GaAs QDs.
The energy levels are compared in three scales: (i) The energy
difference between the lowest electron state and highest hole state, 
which largely determines the exciton energies and is in the order of 1 eV,
(ii) The intraband energy spacing which is in the order of a few tens
meV. 
(iii) The intraband $p$ level splitting due to the C$_{2v}$ atomistic symmetry
of the QDs, which is in the order of a few meV.
The single-particle energy levels of the InAs/InP dots are summarized 
in Table.~\ref{tab:levelspacing},
whereas the results for the InAs/GaAs dots can be found in Table I of 
Ref. \onlinecite{he06a}.  The results of the InAs/GaAs QDs 
are very similar to what was obtained in Ref. \onlinecite{williamson00}.

\subsection{Confined states and wavefunctions} 

Figure \ref{fig:confiedstate}(a) depicts 
the energy levels of all 
confined electron states and 6 highest 
confined hole states of the InAs/InP QDs for 
various sizes. As a reference, we show in 
Fig. \ref{fig:confiedstate}(b) the
lowest 6 confined electron levels and the highest 6 confined hole levels of
a InAs/GaAs QD of $D$=20 nm, $h$=2.5 nm. 
The zero energy is chosen to be the VBM of the host materials. 
The confined electron (hole) states are defined 
to be the states whose energies are lower (higher) than the CBM 
(VBM) of the host materials.
We also show the electron and
hole envelope wavefunctions of the InAs/GaAs and InAs/InP QDs for
two dot geometries: a flat dot ($D$=25 nm, $h$=2.5 nm) in Fig.
\ref{fig:wavefunction} (a) (b) and a tall dot ($D$=25 nm, $h$=5.5
nm) in Fig. \ref{fig:wavefunction} (c) (d). We show the lowest 6
electron and the highest 6 confined hole states for each dot. In all
cases, the isosurface is chosen to enclose 50\% of the total
charge.
The number on the bottom of each small panel gives the
percentage of the density localized inside the QD.

\subsubsection {Confined electron states}

\begin{figure}
\includegraphics[width=3.in,angle=-90]{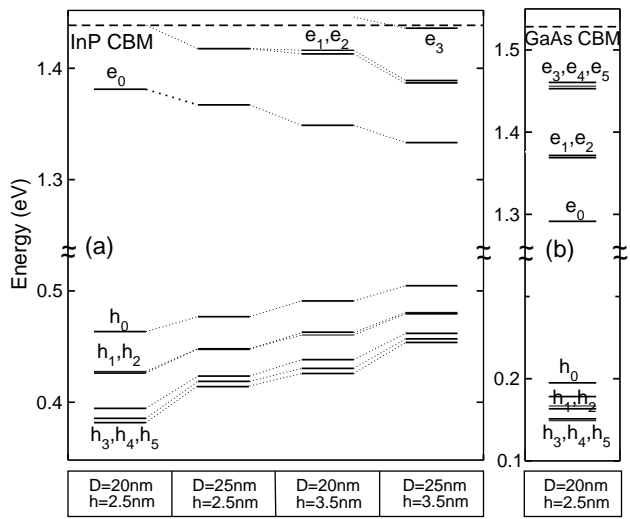}
\caption{Single-particle energy levels for 
(a) the lens-shaped InAs/InP QDs of different sizes, 
and (b) the lens-shaped InAs/GaAs QD ($D$=20 nm, $h$=2.5 nm).
We show all confined electron states and the six highest confined hole states. 
The zero energies are  chosen to be the VBM of the host materials.}
\label{fig:confiedstate}
\end{figure}

As we can see from Fig. \ref{fig:confiedstate}, the InAs/InP QDs
have fewer confined electron states compared to the InAs/GaAs QDs
due to the smaller conduction band offset. In the smaller InAs/InP QD
with $D$=20 nm, $h$=2.5 nm, only one electron state ($e_0$) is
confined. When we increase the dot base and height, more states are confined. 
For the dot with $D$=25 nm, $h$=2.5 nm, $e_0$, $e_1$ and $e_2$  are confined,
whereas in the dot with 
$D$=25 nm, $h$=3.5 nm, the $e_4$ and $e_5$ levels are also confined.
The confined electrons 
show well defined $s$, $p$, $d$ energy level shell structure in both 
the InAs/InP and the InAs/GaAs dots.

\begin{figure*}
\begin{center}
\includegraphics[width=6.0in]{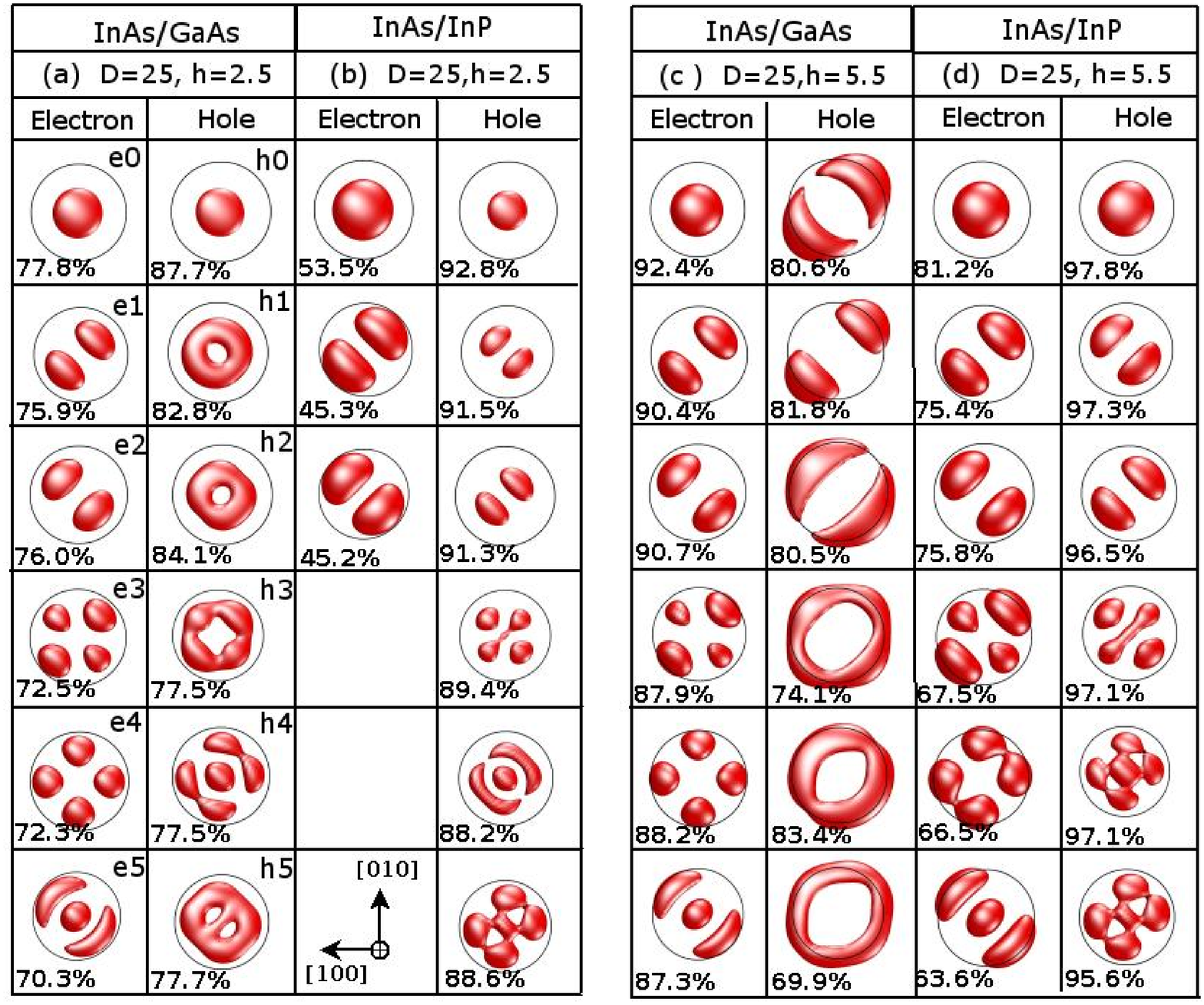}
\end{center}
\caption{(Color online) Top view of the squared
 wavefunctions of the confined electron and hole 
states for the InAs/GaAs and InAs/InP QDs.
We show in (a), (b) the results of flat dots with
 $D$=25 nm,  $h$=2.5 nm,
and in (c) and (d) the results of 
  tall  dots with $D$=25 nm, $h$=5.5 nm. 
  The isosurface to chosen to enclose 50\% of the density. 
The number on the bottom of each small panel gives the
percentage of the density localized inside the QD.
}
\label{fig:wavefunction}
\end{figure*}

For the InAs/GaAs QDs, the electron wavefunctions of the flat dot
[See Fig. \ref{fig:wavefunction} (a)] and the tall dot [See Fig.
\ref{fig:wavefunction} (c)] have very similar shapes. The lowest
state has $s$-like wavefunction, followed by 2 $p$-like states.
The lower energy $p$ orbital ($e_1$) has peaks along the [1$\bar{1}$0] direction,
whereas the higher energy $p$ orbitals ($e_2$) 
has peaks along the
[110] direction.
\cite{orientation}
Considering the next three levels, close in energy, the first 
two orbitals ($e_{4}$, $e_{5}$)
that have similar shapes are $d$ orbitals, 
whereas the third orbital ($e_5$) having a peak in the dot center, is the 
2$s$ orbital.
The wavefunctions of the InAs/InP dots have very similar shapes to the corresponding ones
of the InAs/GaAs dots, 
except that the wavefunctions of the
InAs/InP dots are larger, indicating that the electrons are less
confined in the InAs/InP dots.
In the flat InAs/InP QD, only about 53\% of $e_0$ state is confined in the dot
compared to 78\% in the InAs/GaAs QD of the same size.
Even in the tall InAs/InP QD, less than 82\% of the $e_0$ state is confined
compared to 92\% in the  tall InAs/GaAs QD.

\subsubsection { Confined hole states}
The valence band offset for the InAs/InP QDs is about 530 meV, which is
almost twice as much as in the InAs/GaAs QDs. Therefore, 
more hole states are confined in the InAs/InP dots 
than in the InAs/GaAs dots.
Unlike the InAs/GaAs dot, where the energy level shell structure 
for holes are not so obvious,\cite{he06a}
the InAs/InP dots have well defined hole $s$, $p$, $d$ 
energy level shell structure, similar to those of electrons. 

The shape of the hole wavefunctions in the InAs/GaAs dot is more
complicate than that of the electron wavefunctions. But nevertheless in the
flat dot [See Fig. \ref{fig:wavefunction} (a)] they can still be
recognized as $s$, $p$, $d$, 2$s$ orbitals although with some mixed
characters. \cite{he04a} 
In the tall InAs/GaAs
dot, the holes are strongly localized on the interface of the dot
due to the strain effects, \cite{he04a} and do not have clear $s$,
$p$, $d$ characters any more [See Fig. \ref{fig:wavefunction} (c)].
On the other hand, the hole wavefunctions of the InAs/InP dots are
quite different from those of the InAs/GaAs dots. They are
quite similar to the electron wavefunctions, for both the flat and
tall dots, except that the two $p$ orbitals switch order in
energy, i.e., the first hole $p$ orbital has peaks along
the [110] direction, whereas the second hole $p$ orbital has peaks along
the [1$\bar{1}$0] direction. The rotation of the $p$ orbitals has also been
noticed by Sheng et al.,\cite{sheng05b} and  was attributed to the
piezoelectric effects. However, in our calculations, the piezo-effect
is ignored,
and therefore it cannot be  the reason for the $p$ orbital rotation. 
The rotation of the wavefunctions can be explored experimentally via
magnetocapacitance spectroscopy. \cite{wibbelhoff05}
Unlike in the InAs/GaAs dots, no hole localization has been found in
the InAs/InP dots, because of their smaller strain and the
large confinement potential for holes, 
as is discussed in Sec. \ref{sec:strain}.

The holes are strongly confined in the InAs/InP QDs.  
Even in a small
InAs/InP QD, more than 80\% of charge density is localized in the QD
for the $h_0$ and $h_{1,2}$ states. As a result, in a small InAs/InP QD,
electrons could escape from the dot much more easily than holes,
resulting in a positively charged system. Therefore the InAs/InP dot
might be a good candidate for a memory device via hole storage in
the dot. \cite{pettersson00, verbin06, bipul07}

\subsubsection {Single-particle electron-hole energy gap}
\label{sec:Eeh}

The single-particle electron-hole energy gap
$\Delta \epsilon_{e,h}={e_0}-{h_0}$ is summarized in 
Table \ref{tab:levelspacing}. For the InAs/InP QDs with
$D$=20 nm, and $h$=2.5 nm,  
$\Delta \epsilon_{e,h} \sim$ 917 meV, 
which is about 180 meV less than that of the
InAs/GaAs dot of the same size.
The electron energy levels move down in energy,
whereas the hole energy levels move up in energy, 
with the increasing of the base and height of the dot,
as shown in Fig. \ref{fig:confiedstate} (a).
As a result, $\Delta \epsilon_{e,h}$ decreases with the increasing of the dot
size, due to the reduced confinement. 
For the dot of $D$=20 nm, $\Delta \epsilon_{e,h}$ changes from 918 meV
at $h$=2.5 nm,
to 798 meV at $h$=5.5 nm, with a 120 meV reduction. 
A similar $\Delta \epsilon_{e,h}$  reduction has been found for
the InAs/InP dots of base $D$=25 nm when the dot height increases 
from 2.5 nm to 5.5 nm. 
The electron-hole energy gap reduction
results in a corresponding redshift of the exciton 
emission lines (Sec. \ref{sec:excitons}).


\begin{table}
\caption{Summary of the pseudopotential-calculated single-particle level
  spacing (in meV) of the InAs/InP quantum dots of different base sizes and
  heights (in nm).
Unconfined states are leave in blank in the table.}
\begin{center}
\begin{tabular}{ccccccccccc}\hline \hline
   & \mc{4}{c}{$D$=20}  &\mc{4}{c}{$D$=25} \\
             &$h$=2.5&3.5   &4.5   &5.5   &$h$=2.5&3.5    &4.5   &5.5    \\ \hline
$e_{1}-e_{0}$&56.6 &64.1  &67.4  &67.9  &49.1 &53.6   &56.0  &55.7   \\
$e_{2}-e_{1}$&    &3.2   &4.4   &4.9   &1.4  &2.2    &2.9   &3.4    \\
$e_{3}-e_{2}$&    &      &     &61.1   &     &48.0   &52.6  &53.8   \\
$e_{4}-e_{3}$&    &      &     &       &     &       &2.2   &2.8    \\
$e_{5}-e_{4}$&    &      &     &       &     &       &3.5   &5.3    \\ \hline
$h_{0}-h_{1}$&37.2 &30.6  &27.8  &25.7  &29.6 &25.2   &23.1  &21.3   \\
$h_{1}-h_{2}$&1.5  &2.5   &3.5   &4.2   &0.7  &1.2    &1.5   &1.7   \\
$h_{2}-h_{3}$&31.8 &22.1  &16.1  &13.2  &23.7 &17.6   &14.9  &13.6   \\
$h_{3}-h_{4}$&9.0  &7.9   &6.6   &4.4   &4.8  &4.9    &4.9   &4.2    \\
$h_{4}-h_{5}$&3.8  &4.6   &6.0   &8.2   &4.0  &3.2    &2.9   &3.4    \\ \hline
$e_0-h_0$   &917.7 &857.8 &821.1 &798.4 &890.3 &828.7 &791.1 &768.5  \\ \hline\hline
\end{tabular}
\end{center}
\label{tab:levelspacing}
\end{table}

\subsection{Intraband energy spacings } 

\begin{figure}
\begin{center}
\includegraphics[width=3in,angle=-90]{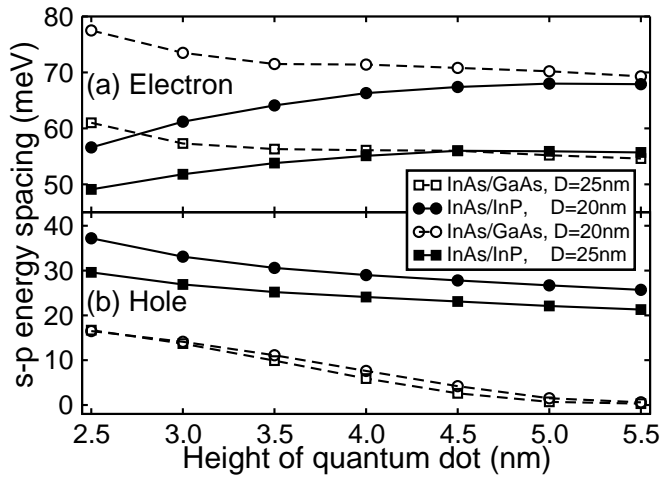}
\end{center}
\caption{Comparison of the intraband energy spacing $\delta_{sp}$ 
between the InAs/InP and InAs/GaAs QDs for 
(a) electrons, and (b) holes. }
\label{fig:spspacing}
\end{figure}

The intraband energy spacing
can be used to characterize the confinement effects in the QDs.
We summarize the $s$-$p$ energy spacing 
$\delta_{sp}={e_1}-{e_0}$ (or $h_0 -h_1$), and 
$p$-$d$ energy level spacing 
$\delta_{pd}={e_3}-{e_2}$ (or $h_2-h_3$) 
of the InAs/InP dots in Table \ref{tab:levelspacing}, 
whereas the results for the InAs/GaAs dots can be found in Table I of 
Ref. \onlinecite{he06a}. 
To further see the trend of how the energy spacing changes with the dot size,
we plot $\delta_{sp}$ of both InAs/GaAs and InAs/InP
QDs, as a function of the dot height in Fig. \ref{fig:spspacing} (a), (b) 
for electrons and holes  respectively.

(a) {\it electrons}:
The electron energy spacing of the InAs/InP dots is in the range of
50 to 70 meV, slightly smaller than the 
energy spacing (50 to 80 meV) of the InAs/GaAs dots,  
due to the weaker confining potential for electrons in the InAs/InP dots.  
The $s$-$p$ energy spacing $\delta_{sp}$ and  $p$-$d$ energy spacing
$\delta_{pd}$ are nearly equal, in rough agreement with the 
EMA with harmonic confinements.

Intuitively, $\delta_{sp}$ should decrease
monotonically by increasing the dot height.  This trend
is followed by the InAs/GaAs dots.
Surprisingly, the electron $\delta_{sp}$ of the InAs/InP dots 
increases with the increasing of the dot height, against the naive expectation.
In this case, both $s$, $p$ levels move down in energy with the increasing 
of the dot size, but 
the $s$ level moves down faster than the $p$ levels, \cite{p-level} 
leading to a larger $\delta_{sp}$.

(b) {\it holes}: The hole energy spacing in the InAs/InP dots ranges from 
20 to 40 meV, significantly larger than that ($<$ 20 meV) of
the InAs/GaAs dots of the same size.
This is because the hole confining potential in the
InAs/InP dots is about 340 meV larger than in the InAs/GaAs dots.  
Nevertheless, in all the InAs/InP QDs we have studied,  
the electron energy spacing is still about twice larger
than that of the holes, 
because electrons have a much lighter effective mass than holes.
For holes, $\delta_{pd}$ is much smaller than $\delta_{sp}$,
deviating from the harmonic potential approximation. 
As shown in Fig. \ref{fig:spspacing} (b),
the energy spacing $\delta_{sp}$
of holes decrease monotonically
with the increasing of the dot height for both InAs/GaAs
and InAs/InP QDs.  No anomaly is found.
Notice that for very tall QDs, $\delta_{sp}$ 
becomes very small for the InAs/GaAs QDs, due to the hole localization on
the interface of the QDs [see
Fig. \ref{fig:wavefunction} (c)], which is not the case for the InAs/InP QDs.

\subsection {Intraband $p$ level splitting}

In a continuum model, ignoring the underlying atomistic structure, 
a cylindrical QD has the $C_{\infty v}$ symmetry, leading to
degenerate $p$, $d$ levels.  
In contrast, the atomistic theories 
maintain the real symmetry of the QDs.
For example, the lens-shaped QDs made of zinc-blende 
III-V semiconductors are of 
$C_{2v}$ symmetry,\cite{bester05a} where the
[110] and [1$\bar{1}$0] directions are non-equivalent, 
resulting in split $p$ levels and $d$ 
levels. 
The values of the $p$-$p$ splitting 
$\delta_{pp}= {e_2}- {e_1}$ (or ${h_1}- {h_2}$) 
are summarized in Table \ref{tab:levelspacing} for the InAs/InP QDs.
We also depict the $p$-$p$ splittings as functions of dot height
for both InAs/InP and InAs/GaAs QDs in Fig. \ref{fig:psplitting}. 

\subsubsection { $p$ level splitting for electrons}

\begin{figure}
\begin{center}
\includegraphics[width=2.7in, angle=-90]{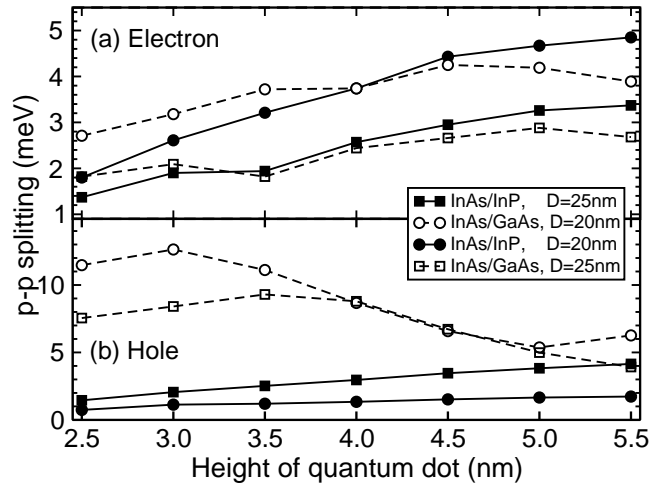}
\end{center}
\caption{ Comparison of the $p$ level splitting $\delta_{pp}$ 
between the InAs/InP and InAs/GaAs QDs for 
(a) electrons, and (b) holes.
}
\label{fig:psplitting}
\end{figure}

The electron $p$-$p$ splittings are shown in Fig. \ref{fig:psplitting} (a)
as functions of the dot height. 
The electron $\delta_{pp}$ is $\sim$ 1 - 2 meV for
flat InAs/InP dot, and increases to about 3 - 5 meV for the tall dots. 
The dots with the smaller base ($D$=20 nm) show a much larger $p$-$p$ splitting. 
The electron $p$-$p$ splitting for the InAs/InP dots
is about 3\%-5\% of the electron $s$-$p$ energy spacing $\delta_{sp}$.  
The electron $p$-$p$ splittings for the InAs/GaAs dots are in the range of 2 -
4 meV, and 3 \% - 4\% of $\delta_{sp}$, and show a weak dependence on the dot height. 

\subsubsection{$p$ level splitting for holes}

The hole $p$-$p$ splittings are shown
in Fig. \ref{fig:psplitting} (b).
It has been shown in previous studies, \cite{williamson00,he05d,he06a} that
the hole $p$-$p$ splittings can be larger than 10 meV
in the InAs/GaAs dots, which is about 70\% of the $s$-$p$ energy
spacing $\delta_{sp}$. 
For very tall InAs/GaAs dots, in which the holes
localize on the interface,  $\delta_{pp}$ could even be
much larger than $\delta_{sp}$. 
The large $p$-$p$ splitting leads to very different
electronic and optical properties of the InAs/GaAs QDs
than those predicted by continuum theories,
e.g. the nontrivial charging pattern that breaks Hund's rule and the Aufbau
principle for holes,\cite{he05d,he06a} that has been recently confirmed 
experimentally. \cite{reuter05, bester07}
Surprisingly, the calculated hole $p$-$p$ splitting of the InAs/InP dot is 
much smaller than that of the InAs/GaAs dots, even though they have the
same dot materials.
For the flat InAs/InP dots, 
the splitting is only about 1 $\sim$ 2 meV and for the tall
dot with a small base ($D$ =20 nm) the splitting can be as large as 4.2 meV. 
Nevertheless, the hole $\delta_{pp}$  
is less than 10 \% of $\delta_{sp}$ for the InAs/InP dots.
Therefore the multi-hole phase diagram and the charging patterns in the
InAs/InP dots
are expected to be very different from those in the InAs/GaAs dots, which can be 
examined by the hole charging experiments. \cite{reuter05}
This is one of the most important results of the present work, which can not
be obtained from the continuum theories. For example, a k$\cdot$p theory 
predicted small ($\sim$ 1 meV) 
hole $p$-$p$ splitting for both types of dots. \cite{lee04,sheng05, sheng05b}
The difference in the hole splitting for the two types of dots might comes from their
different strain profiles, the band offset, or the interface effects (common
cation vs. common anion). We leave this for future investigations. 

\begin{figure}
\begin{center}
\includegraphics[width=2.6 in,angle=-90]{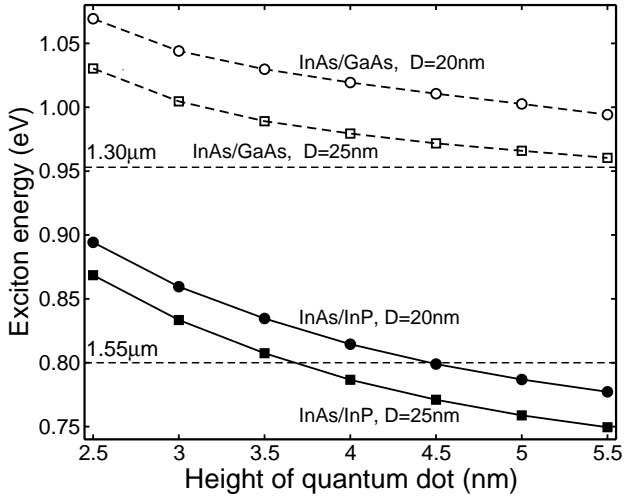}
\end{center}
\caption{ Comparison of the primary exciton energies vs. dot height between the
  InAs/InP and InAs/GaAs QDs. }
\label{fig:exciton}
\end{figure}

\begin{figure}
\begin{center}
\includegraphics[width=2.8in, angle=-90]{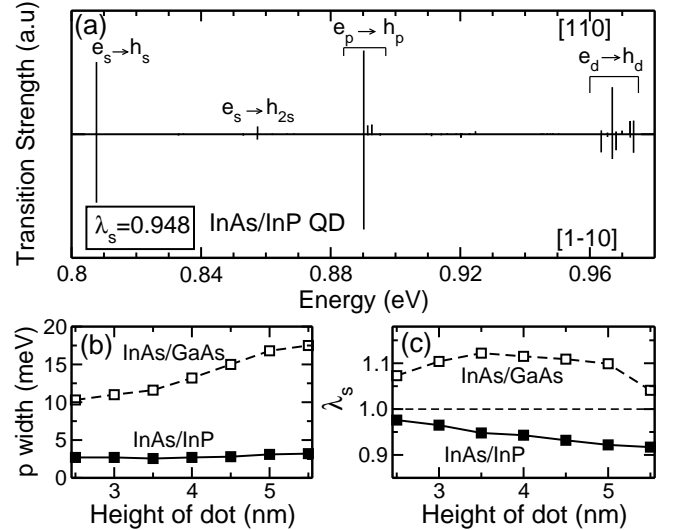}
\end{center}
\caption{(a) The exciton transition strengths for the InAs/InP QD of
$D$=25 nm, $h$=3.5 nm. 
The upper panel is for the transitions polarized in the [110] direction, 
while the lower panel is for transitions polarized in the [1$\bar{1}$0]
direction.  
$\lambda_s$ is the polarization anisotropy (Eq. \ref{eq:lambda}) 
of the $S$ exciton.
(b) The exciton $P$ shell width vs. dot height. 
(c) The $S$ exciton polarization anisotropy
$\lambda_s$ vs. dot height.
}
\label{fig:Xdipole}
\end{figure}

\section{Excitons}        
\label{sec:excitons}

Figure \ref{fig:exciton} depicts the fundamental exciton energies 
vs. dot height for the InAs/InP and InAs/GaAs dots. 
The exciton energies of the InAs/GaAs dots are about 200 meV higher than 
those of the InAs/InP dots.
The exciton energies 
decrease monotonically with increasing of the dot size.  
For the InAs/InP dot with $D$=20 (25) nm, the exciton energy reduces
from about 900 (870) meV to about 780 (750) meV, 
as the dot height increases from 2.5 nm
to 5.5 nm. The exciton energy 
reduction is about 120 meV, compared to 70 meV for the InAs/GaAs dots.
One of the most important motivations of studying the InAs/InP dots is the
1.55 $\mu$m ($\sim$ 800 meV) emission for device applications. 
This wavelength can be easily achieved by the InAs/InP
dots with a reasonable size, 
while it is challenging to be obtained using the InAs/GaAs dots,
as shown in Fig. \ref{fig:exciton}. 

In the Hartree-Fock approximation, the fundamental exciton
energy can be calculated as,
\begin{equation}
E_X=\Delta \epsilon_{e,h} -J^{(eh)} \, ,
\end{equation}
where $\Delta \epsilon_{e,h}={e_0}-{h_0}$ is 
the single-particle electron-hole
energy gap and
$J^{(eh)}$ is the direct electron-hole
Coulomb energy. The electron-hole exchange energy
is much smaller than $J^{(eh)}$ and therefore is ignored here.
In typical InAs dots, $\Delta \epsilon_{e,h} \gg J^{(eh)}$.
Therefore the exciton energy is largely determined by the single-particle electron-hole
energy gap $\Delta \epsilon_{e,h}$ (See Sec: \ref{sec:Eeh}).

We further calculated the higher excitonic transitions for the InAs/InP dot 
with $D$=25 nm and $h$=3.5 nm, for which the fundamental exciton wavelength 
is close to 1.55 $\mu$m. 
The results are shown in Fig. \ref{fig:Xdipole} (a), where the upper panel 
depicts the transition strength along the [110] direction, while the lower panel
shows the transition strength along the [$1\bar{1}0$] direction.  
The transitions polarized along the [001] direction are about 4-5 orders of magnitude
smaller than those parallel to the (001) plane, 
and are therefore not considered here.  
The excitonic transitions form several shells: 
the first shell coming from the $e_0$ to $h_0$ transitions is the exciton $S$ shell.
The transitions of $e_{1,2}$ to $h_{1,2}$, form the exciton $P$ shell, 
whereas the transitions of $e_{3,4,5}$ to $h_{3,4,5}$ form the $D$ shell. 
\cite{bayer00b, hawrylak00}
There is also a small transition peak at about 0.857 eV, which comes from 
the recombination of $e_0$ electron and $h_5$ hole. 
This transition, however, is significantly weaker in the InAs/GaAs QDs.
The exciton $P$ shell and $D$ shell are composed of a bunch of transitions with
slightly different transition energies.
For the InAs/GaAs QDs, the width of the $P$ shell is about 12 meV, whereas
for the InAs/InP QD, the width is only about 2.5 meV.
The $D$ shell width is about 10 meV for the InAs/InP QDs, compared
to about 22 meV in the InAs/GaAs dots.  
The $P$ shell width with respect to the height of the QDs is 
presented in Fig. \ref{fig:Xdipole} (b) for the dot with $D$=25 nm. 
The $P$ shell width of the InAs/GaAs increases to about 20 meV for $h=$5.5
nm. However, for InAs/InP dots, we
found that the width is always about 2 $\sim$ 3 meV. 
This feature reflects the fact that the $p$ level splitting is small 
for both electrons and holes in the
InAs/InP dots, fact that could be examined 
by the single-dot optical spectroscopy. \cite{kuther98, hawrylak00}

We also calculated the light polarization anisotropy 
$\lambda$, 
defined as the ratio of the transition intensities along
the [110] and $[1\bar{1}$0] direction, i.e., \cite{williamson00}
\begin{equation}
\lambda ={I_{[110]} \over I_{[1\bar{1}0]}} \, .
\label{eq:lambda}
\end{equation}
The results for the $S$ shell excitons in the dot with base $D$=25 nm
are presented in Fig. \ref{fig:Xdipole} (c) for both the InAs/InP and
InAs/GaAs dots.
For the InAs/GaAs dot, we found $\lambda_s >$ 1, indicating that the 
intensity along the $[110]$ direction is stronger than 
that along the [1$\bar{1}$0] direction, \cite{williamson00}
whereas, in the InAs/InP QD, $\lambda_s <$ 1, indicating that the stronger intensity 
is along $[1\bar{1}0]$ direction. 
This feature could also be examined by the optical spectroscopy.

\section{Conclusion}
\label{sec:conclusion}

We have studied the electronic structure of the 
InAs/InP QDs using an atomistic pseudopotential method
and compared them to those of the
InAs/GaAs QDs. 
Our results show that 
even though the InAs/InP QDs and InAs/GaAs QDs have the same dot material,
their electronic structure and optical properties
differ significantly 
in certain aspects.
These features, which may have important
impacts for device applications,
could be examined in future experiments. 
Some of the features can only be captured by atomistic theories
and therefore provide a unique opportunity to test the 
predictive capability of the different theoretical
approaches.

\begin{table}
\caption{Band parameters obtained from the pseudopotential band structure and the target
values of the fit.
The ``Target values'' are conventional bulk parameters used in the literature 
(see Ref.\onlinecite{vurgaftman01}).
$\Delta E_{VBO}$ is the valence band offset relative
to the  bulk InAs VBM. $\Delta_0$ is the spin-orbit splitting, $m*$  are the
effective masses at $\Gamma$, and $a_g$, $a_v$ and $b$ are the hydrostatic deformation
potentials of the band gap, the valence band maximum, and the biaxial deformation
potential of the valence band, respectively. The predicted band structure critical points are
compared with the existing experimental data\cite{a28}.
}
\begin{center}
\begin{tabular}{ccccc} \hline \hline
\multicolumn{1}{c}{Parameters}
& \multicolumn{2}{c}{InAs} & \multicolumn{2}{c}{InP} \\
\hline
    & PP & Target & PP & Target  \\
\hline
\multicolumn{5}{c}{Fit}\\
\hline
$E_g$ (eV) & 0.410 & 0.410 & 1.424 & 1.424  \\
$\Delta E_{VBO}$ (eV) &  -0.006 & 0.000 & -0.440 & -0.420  \\
$\Delta _0$ (eV)  & 0.390 & 0.390 & 0.109  & 0.108   \\
$m^{*}_e$  & 0.022 & 0.024 & 0.059 & 0.080  \\
$m^{*}_{hh}(001)$ & 0.387 & 0.341 & 0.444  &  0.520 \\
$m^{*}_{hh}(111)$ & 1.006 & 0.917 & 1.180 & 0.950  \\
$m^{*}_{lh}(001)$ & 0.027 & 0.027 & 0.085 & 0.110  \\
$m^{*}_{lh}(111)$ & 0.026 & 0.026 & -  & - \\
$m^{*}_{so}(001)$ & 0.097 & 0.085 & 0.152  & 0.21  \\
$a_g$  & -6.44 & -6.6 & -6.93  & -6.0  \\
$a_v$  & -1.01 & -1.0 & -0.68 & -0.6  \\
$b$    & -1.78 & -1.70 & -1.67 & -2.0  \\
\hline
\multicolumn{5}{c}{Predictions} \\
\hline
$\Gamma_{7c}$ (eV) & 4.55 &  4.52  &  5.31 & 4.72 \\
$X_{6v}$  & -2.38 &  -2.4 & -2.38 &  -2.3 \\
$X_{7v}$  & -2.37 &  -2.4 & -2.25 & -2.2 \\
$X_{6c}$  & 2.28 &   - &  2.21 &  2.38 \\
$X_{7c}$  &  2.29  &  -   & 2.61 &  - \\
$L_{6v}$  & -1.14  & -0.90 & -0.92 & -1.23 \\
$L_{4,5v}$ & -0.87 & -0.90  & -0.80 & -1.12 \\
$L_{6c}$ &  1.46 &  - &  2.15 & 2.03 \\
 \hline\hline
\end{tabular}
\end{center}
\label{tab:fitproperties}
\end{table}

\appendix

\section{EPM for InAs/InP}
\label{sec:EPM}

As explained in Sec. II the crystal potential is written as a superposition of atomic 
potentials $v_{\alpha}$
centered on the atomic positions. For each atomic potential
we use for the screened pseudopotentials the expression proposed by Williamson et al.\cite{williamson00}
\begin{widetext}
\begin{equation}
V_{\alpha}(r-R_{n\alpha})= v_{\alpha}(r-R_{n\alpha}) \left[ 1 + \delta v_{n\alpha}(\epsilon) \right] 
={1 \over \omega_c} \left( \sum_{\mathbf{q}} e^{i\mathbf{q} \cdot (r-R_{n\alpha})} v_{\alpha}(|\mathbf{q}|) \right)
\left[ 1 + \delta v_{n\alpha}(\epsilon) \right], 
\end{equation}
\end{widetext}
where $v_{\alpha}(|\mathbf{q}|)$ has the functional form:
\begin{equation}
v_{\alpha}(|{\mathbf{q}}|) = a_{0\alpha} \cdot { {q^2 - a_{1\alpha}} \over { a_{2\alpha} e^{a_{3\alpha}q^2} - 1 } }, \label{appb}
\end{equation}
and
\begin{equation}
\delta v_{n\alpha}(\epsilon) = \gamma_{\alpha} \cdot (\epsilon_{xx}+\epsilon_{yy}+\epsilon_{zz}).
\end{equation}
$\epsilon_{ii}$ are elements of the local strain tensor.
The term $\delta v_{n\alpha}(\epsilon)$ plays a crucial role in describing the
absolute hydrostatic deformation potentials, in particular the variation of the valence band edge and, separatly, the
conduction band edge under arbitrary strains. This allows us to describe the modification
of the valence and conduction band offsets when the systems are subjected to hydrostatic
or biaxial deformation conditions such as in the case of epitaxial growth on a lattice-mismatched
substrate.
The parameters entering the previous equations have been determined 
by fitting a number of experimentally and theoretically (ab-initio) determined properties of
bulk InP and InAs:  the
experimentally measured electron and hole effective masses, band gaps (target values at 0$^{\circ}$ K),
and spin-orbit splittings,
hydrostatic deformation potentials of the band gaps, band offsets, and LDA-predicted single
band edge deformation potentials\cite{wei98}.
In the previous equation the term $\beta$ has been introduced to represent the quasiparticle nonlocal
self-energy effects. This kinetic energy scaling is needed to simultaneously fit
bulk effective masses and band gaps.

\begin{table}
\caption{Fitted pseudopotential parameters for InAs/InP. 
A plane-wave cutoff of 5 Ryd is used. 
}
\label{tab:empricalparameters}
\begin{tabular}{llllllll} \hline \hline
parameters  & As(In)   &  P(In)   & In(As)  & In(P) \\ \hline
$\alpha_0$      &56.8819 &0.1509   &853.4653  &5012.0545  \\ 
$\alpha_1$      &2.7023  &2.9215   &1.9724    &1.8556     \\
$\alpha_2$      &1.4894  &1.2190   &19.1236   &88.8570    \\
$\alpha_3$      &0.5757  &0.3554   &0.5439    &0.7419     \\
$\gamma_{\alpha}$&0.00   &0.00     &1.6597    &1.6460     \\
$\alpha_{so}$   &0.1315  &0.0140   &0.4056    &0.4800     \\ 
\hline \hline
\end{tabular}
\end{table}

In  Table.~\ref{tab:fitproperties} we report the target values we have fit for the  binary 
InAs, and InP, and the results of the fitting procedure. 
The target values correspond to the
band parameters used in the literature\cite{vurgaftman01} at T = 0K. 
A 5 Ry kinetic cutoff was used when generating the
pseudopotentials. This cutoff has then been used in the QD calculations 
of this paper. From Table \ref{tab:fitproperties} we see that the fit is satisfying.
The corresponding parameters of the empirical
pseudopotentials  are given in Table \ref{tab:empricalparameters}. Although we fitted only a few band
properties per material, we checked that the fit works also for the full band structure.
The predicted (not fitted) critical point energies are also reported in Table \ref{tab:fitproperties}.
Notice that we are using slightly different potentials for the In atoms in InP and InAs, to take
into account the different charge redistribution occurring around the In atom when it is placed in
a different environment.
Considering only the nearest-neighbor environment, the potential of each In atom in the
structure is obtained as:
\begin{equation}
v_{{\rm In}}({\rm As}_n{\rm P}_{4-n}) = {n \over 4} v_{{\rm In}}({\rm InAs}) + { {4-n} \over 4} v_{{\rm In}}({\rm InP})
\end{equation}

\acknowledgments
L.H. acknowledges the support from the Chinese National
Fundamental Research Program, the Innovation
funds and ``Hundreds of Talents'' program from Chinese Academy of
Sciences, and National Natural Science Foundation of China (Grant
No. 10674124).
R. M. acknowledges support from the italian MIUR PRIN 2005.


\begin{thebibliography}{50}
\bibitem[{ori()}]{orientation}
\bibinfo{note}{We adopt the definiation of the crystallographic direction of
  Ref. \onlinecite{bester05a}, which choose to place the cation at the origin
  (0,0,0,). In Ref. \onlinecite{williamson00}, a different convention for the
  [110] direction has been used.}
\bibitem[{p-l()}]{p-level}
\bibinfo{note}{The $p$ levels are slightly higher than the VBM of InP, and
  therefore are weakly confined when dot height $h <$ 3.5 nm.}
\end{thebibliography}
\end{document}